\documentclass[useAMS,usenatbib]{mn2e}

\usepackage{graphicx}
\usepackage{amsmath}
\usepackage{amssymb}
\usepackage{color}
\usepackage{subfig}
\usepackage[breaklinks,colorlinks,citecolor=blue,linkcolor=red]{hyperref} 
\usepackage[all]{hypcap}

\newcommand\msun{\, \rm M_\odot}

\newcommand\kms{\, \rm km\,s^{-1}}

\newcommand\gyr{{\, \rm Gyr}}

\newcommand\eout{{e_{\rm out}}}
\newcommand\ein{{e_{\rm in}}}
\newcommand\aout{{a_{\rm out}}}
\newcommand\ain{{a_{\rm in}}}
\newcommand\amax{{a_{\rm 3, max}}}
\newcommand\mbh{{m_{\rm BH}}}
\newcommand\mns{{m_{\rm NS}}}
\newcommand\sigbh{{\sigma_{\rm BH}}}
\newcommand\signs{{\sigma_{\rm NS}}}

\newcommand\vk{{v_\mathrm{k}}}

\title[BH-NS mergers from triples]{Black hole-neutron star mergers from triples}
\author[G. Fragione, A. Loeb]{\parbox{\textwidth}{Giacomo Fragione$^{1}$\thanks{E-mail: giacomo.fragione@mail.huji.ac.il}, Abraham Loeb$^{2}$}\\
$^1$Racah Institute for Physics, The Hebrew University, Jerusalem 91904, Israel\\
$^2$Astronomy Department, Harvard University, 60 Garden St., Cambridge, MA 02138, USA}

\begin{document}

\maketitle

\begin{abstract}
Mergers of black hole (BH) and neutron star (NS) binaries are expected to be observed by gravitational wave observatories in the coming years. Until now, LIGO has only set an upper limit on this merger rate. BH-NS binaries are expected to merge in isolation, as their formation is suppressed in star clusters by the mass segregation and the strong heating by BHs in the cluster core. Another viable scenario to BH-NS mergers is in triple systems. In this paper, we carry out a systematic statistical study of the dynamical evolution of triples comprised of an inner BH-NS binary by means of high-precision $N$-body simulations, including Post-Newtonian (PN) terms up to 2.5PN order. We start from the main sequence massive stars and model the supernovae (SN) events that lead to the formation of BHs and NSs. We adopt different prescriptions for the natal velocity kicks imparted during the SN processes and illustrate that large kicks lead to more compact and massive triples that merge on shorter timescales. We also show that BH-NS merging in triples have a significant eccentricity in the LIGO band, typically much larger than BH-NS merging in isolated binaries. Finally, we estimate a rate of $\Gamma_\mathrm{BH-NS}\approx 1.0\times 10^{-3}-3.5\times 10^{-2} \ \mathrm{Gpc}^{-3}\ \mathrm{yr}^{-1}$, for non-zero velocity kicks, and $\Gamma_\mathrm{BH-NS}=19 \ \mathrm{Gpc}^{-3}\ \mathrm{yr}^{-1}$, for no natal kicks. Our rate estimate overlaps with the expected BH-NS rate in isolated binaries and within the LIGO upper limit.
\end{abstract}

\begin{keywords}
galaxies: kinematics and dynamics -- stars: black holes -- stars: neutron -- stars: kinematics and dynamics -- Galaxy: kinematics and dynamics
\end{keywords}

\section{Introduction}

The LIGO-Virgo collaboration has recently released a catalogue of compact object mergers observed via gravitational wave (GW) emission. The catalogue includes ten black hole (BH) binaries and one neutron star (NS) binary \citep{ligo2018}. From the first two observing runs, the LIGO-Virgo collaboration inferred a merger rate of $110$--$3840$ Gpc$^{-3}$ yr$^{-1}$ for binary NSs and a merger rate of $9.7$--$101$ Gpc$^{-3}$ yr$^{-1}$ for binary BHs. As the detector sensitivity improves and new instruments are developed, hundreds of merging binary signals are expected to be observed in the upcoming years, and the modeling of astrophysical channels that lead to the formation of BH and NS binaries has become of crucial importance.

Several different astrophysical channels have been proposed to form merging compact objects. Possibilities include isolated binary evolution through a common envelope phase \citep{bel16b,giac2018,kruc2018} or chemically homogeneous evolution \citep{mand16,march16}, GW capture events in galactic nuclei \citep{olea09,rass2019}, mergers in star clusters \citep{askar17,baner18,frak18,rod18}, Kozai-Lidov (KL) mergers of binaries in galactic nuclei \citep{antoper12,petr17,fragrish2018,grish18}, in stellar triple \citep{ant17,sil17,arc2018} and quadruple systems \citep{fragk2019,liu2019}, and mergers in active galactic nuclei accretion disks \citep{bart17,sto17}. While typically each model accounts for roughly the same rate of the order of a $\sim\ \mathrm{few}$ Gpc$^{-3}$ yr$^{-1}$, the statistical contribution of different astrophysical channels can be hopefully disentangled using the distributions of their masses, spins, eccentricity and redshift \citep[see e.g.][]{olea16,gondan2018,zevin18}.

Generally, theoretical predictions can match the rate inferred by LIGO for BHs, while the high rate estimated for NS-NS inferred from GW170817 can be explained only by assuming rather extreme prescriptions about natal kicks and common envelope in binary systems. Interestingly, BH-NS mergers have not been observed as of yet and LIGO has only set a $90\%$ upper limit of $610$ Gpc$^{-3}$ yr$^{-1}$ on the merger rate. The formation of such binaries is intriguing. The most natural is in isolation \citep{giac2018,kruc2018}. In star clusters, mass segregation and the strong heating by gravitational BH scattering prevent the NSs from forming BH-NS binaries \citep{frag2018}. As a result, BH-NS binaries are expected to be even rarer than NS-NS binaries in star clusters. Recent simulations by \citet{ye2019} have shown that only a very few BH-NS binaries are formed even in massive clusters. Another viable scenario is the dynamical formation in galactic nuclei, but aside from \citet{fragrish2018} this scenario was not studied in detail.

Bound stellar multiples are common. Surveys of massive stars, which are the progenitors of NSs and BHs, have shown that more than $\sim 50$\% and $\sim 15$\% have at least one or two stellar companions, respectively \citep{duq91,ragh10,sa2013AA,duns2015,sana2017,jim2019}. Most previous studies on bound multiples have exclusively focused on determining the BH-BH merger rate from isolated bound triples \citep{ant17,sil17} or quadruples \citep{fragk2019,liu2019}.

In this paper, we study for the first time the dynamical evolution of triples comprised of an inner BH-NS binary by means of high-precision $N$-body simulations, including Post-Newtonian (PN) terms up to 2.5PN order and GW emission. We denote the BH and NS masses in the inner binary as $\mbh$ and $\mns$, respectively, and the mass of third companion as $m_3$. We start from the main-sequence (MS) progenitors of the compact objects and model the supernova (SN) events that lead to the formation of BHs and NSs. We adopt different prescriptions for the natal kick velocities that are imparted by SN events. We quantify how the probability of merger depends on the initial conditions and determine the parameter distribution of merging systems relative to the initial distributions, showing that BH-NS mergers in triples predict a high merger rate only in the case low or null natal velocity kicks are assumed. Most of these mergers have high eccentricity ($\gtrsim 0.1$) in the LIGO frequency band.

The paper is organized as follows. In Section~\ref{sect:supern}, we discuss the SN mechanism in triple stars. In Section~\ref{sect:lk}, we discuss the KL mechanism in triple systems. In Section~\ref{sect:results}, we present our numerical methods to determine the rate of BH-NS mergers in triples, and discuss the parameters of merging systems. Finally, in Section~\ref{sect:conc}, we discuss the implications of our findings and draw our conclusions.

\section{Supernovae in triples}
\label{sect:supern}

\begin{table*}
\caption{Models parameters: name, dispersion of BH kick-velocity distribution ($\sigbh$), dispersion of the NS kick-velocity distribution ($\signs$), eccentricity distribution ($f(e)$), maximum outer semi-major axis of the triple ($\amax$), fraction of stable triple systems after SNe ($f_{\rm stable}$), fraction of stable systems that merge from the $N$-body simulations ($f_{\rm merge}$).}
\centering
\begin{tabular}{lcccccc}
\hline
Name & $\sigbh$ ($\kms$) & $\signs$ ($\kms$) & $f(e)$ & $\amax$ (AU) & $f_{\rm merge}$ & $f_{\rm stable}$\\
\hline\hline
A1 & $\signs\times m_\mathrm{BH}/m_\mathrm{NS}$ & $260$ & uniform  & $2000$ & $1.1\times 10^{-6}$ & $0.11$\\
A2 & $0$  & $0$ & uniform  & $2000$ & $1.6\times 10^{-2}$ & $0.08$\\
A3 & $\signs\times m_\mathrm{BH}/m_\mathrm{NS}$ & $100$ & uniform  & $2000$ & $3.6\times 10^{-5}$ & $0.11$\\
B1 & $\signs\times m_\mathrm{BH}/m_\mathrm{NS}$ & $260$ & thermal  & $2000$ & $1.6\times 10^{-6}$ & $0.09$\\
C1 & $\signs\times m_\mathrm{BH}/m_\mathrm{NS}$ & $260$ & uniform  & $3000$ & $0.9\times 10^{-6}$ & $0.10$\\
C2 & $\signs\times m_\mathrm{BH}/m_\mathrm{NS}$ & $260$ & uniform  & $5000$ & $0.7\times 10^{-6}$ & $0.10$\\
\hline
\end{tabular}
\label{tab:models}
\end{table*}

We consider a hierarchical triple system that consists of an inner binary of mass $m_{\rm in}=m_1+m_2$ and a third body of mass $m_3$ that orbits the inner binary \citep{pijloo2012,toonen2016,lu2019}. The triple can be described in terms of the Keplerian elements of the inner orbit, describing the relative motion of $m_1$ and $m_2$, and of the outer orbit, describing the relative motion of $m_3$ around the centre of mass of the inner binary. The semi-major axis and eccentricity of the inner orbit are $\ain$ and $\ein$, respectively, while the semi-major axis and eccentricity of the outer orbit are $\aout$ and $\eout$, respectively. The inner and outer orbital plane have initial mutual inclination $i_0$.

When the primary star undergoes an SN event, we assume that it takes place instantaneously, i.e. on a time-scale shorter than the orbital period, during which the star has an instantaneous removal of mass. Under this assumption, the position of the body that undergoes an impulsive SN event is assumed not to change. We ignore the SN-shell impact on the companion stars. As a consequence of the mass loss, the exploding star is imparted a kick to its center of mass \citep{bla1961}. Moreover, the system receives a natal kick due to recoil from an asymmetric supernova explosion. We assume that the velocity kick is drawn from a Maxwellian distribution
\begin{equation}
p(\vk)\propto \vk^2 e^{-\vk^2/\sigma^2}\ ,
\label{eqn:vkick}
\end{equation}
with a velocity dispersion $\sigma$.

We consider first the SN event of the primary star in the inner binary. Before the SN takes place, the inner binary consists of two stars with masses $m_1$ and $m_2$, respectively, a relative velocity, $v=|{\bf{v}}|$, and a separation distance, $r=|{\bf{r}}|$. Conservation of the orbital energy implies
\begin{equation}
|{\bf{v}}|^2=\mu\left(\frac{2}{r}-\frac{1}{\ain}\right)\ ,
\label{eqn:vcons}
\end{equation}
with $\mu=G(m_1+m_2)$, while the specific relative angular momentum $\bf{h}$ is related to the orbital parameters as
\begin{equation}
|{\bf{h}}|^2=|{\bf{r}}\times {\bf{v}}|^2=\mu \ain(1-e_{\mathrm{in}}^2)\ .
\label{eqn:hcons}
\end{equation}
After the SN event, the orbital semi-major axis and eccentricity change as a consequence of the mass loss $\Delta m$ and a natal kick velocity $\vk$. The total mass of the binary decreases to $m_{1,n}+m_2$, where $m_{1,n}=m_1-\Delta m$, while the relative velocity will become ${\bf{v_n}}={\bf{v}}+{\bf{\vk}}$. Since ${\bf{r_{\rm n}}}={\bf{r}}$, the new semi-major axis can be computed from Eq.~\ref{eqn:vcons},
\begin{equation}
a_{\rm in,n}=\left(\frac{2}{r}-\frac{v_n^2}{\mu_{\rm in,n}}\right)^{-1}\ ,
\end{equation}
where $\mu_{\rm in,n}=G(m_{1,n}+m_2)$, while the new eccentricity can be computed from Eq.~\ref{eqn:hcons},
\begin{equation}
e_{\rm in,n}=\left(1-\frac{|{\bf{r}}\times {\bf{v_n}}|^2}{\mu_{\rm in,n} a_{\rm in,n}}\right)^{1/2}\ .
\end{equation}

Due to the SN of the primary in the inner binary, an effective kick ${\bf{V_{\rm cm}}}$ is imparted to its center of mass. As a consequence, its position ${\bf{R_0}}$ changes by ${\bf{\Delta R}}$ to ${\bf{R}}={\bf{R_0}}+{\bf{\Delta R}}$, and the outer semi-major axis $\aout$ and eccentricity $\eout$ change accordingly. To compute the relative change with respect to the pre-SN values, the strategy is to use again Eq.~\ref{eqn:vcons} and Eq.~\ref{eqn:hcons} relative to the outer orbit and the outer orbit relative velocity ${\bf{V_{\rm 3}}}$, with effective velocity kick ${\bf{V_{\rm cm}}}$,
\begin{equation}
a_{\rm out,n}=\left(\frac{2}{R}-\frac{V_n^2}{\mu_{\rm out,n}}\right)^{-1}\ ,
\end{equation}
where $\mu_{\rm out,n}=G(m_{1,n}+m_2+m_3)$ and ${\bf{V_{\rm n}}}={\bf{V_{\rm 3}}}+{\bf{V_{\rm cm}}}$,
\begin{equation}
e_{\rm out,n}=\left(1-\frac{|{\bf{R}}\times {\bf{V_n}}|^2}{\mu_{\rm out,n} a_{\rm out,n}}\right)^{1/2}\ .
\end{equation}

Finally, the inclination of the outer binary orbital plane with respect to the inner binary orbital plane is tilted as a consequence of the kick. The new relative inclination $i_{\rm n}$ can be computed from,
\begin{equation}
\sin i_{\rm n}=\frac{|{\bf{R_0}}|}{|{\bf{R}}|} \sin i_0\ .
\end{equation}

The same prescriptions can be applied for the other two stars in the triple. After every SN event, if either $\ain\le 0$ or $\aout\le 0$ the triple is considered to have had a merger, while if either $\ein\ge 1$ or $\eout\ge 1$ the triple is considered to be unbound.

\section{Kozai-Lidov mechanism}
\label{sect:lk}

We consider triple objects where the inner binary is made up of a BH and a NS, of mass $\mbh$ and $\mns$, respectively. A triple system undergoes KL oscillations in eccentricity whenever the initial mutual orbital inclination of the inner and outer orbit is in the window $i\sim 40^\circ$-$140^\circ$ \citep{koz62,lid62}. At the quadrupole order of approximation, the KL oscillations occur on a timescale \citep{nao16},
\begin{equation}
T_{\rm KL}=\frac{8}{15\pi}\frac{m_{\rm tot}}{m_{\rm 3}}\frac{P_{\rm out,n}^2}{P_{\rm BHNS}}\left(1-e_{\rm out,n}^2\right)^{3/2}\ ,
\end{equation}
where $m_{\rm 3}$ is the mass of the outer body orbiting the inner BH-NS binary, $m_{\rm tot}$ is the total mass of the triple system, and $P_{{\rm BHNS}}\propto a_{{\rm in,n}}^{3/2}$ and $P_{{\rm out,n}}\propto a_{{\rm out,n}}^{3/2}$ are the orbital periods of the inner BH-NS binary and of the outer binary, respectively. The maximal eccentricity is essentially a function of the initial mutual inclination
\begin{equation}
e_{\rm in,n}^{\max}=\sqrt{1-\frac{5}{3}\cos^2 i_\mathrm{n}}\ .
\label{eqn:emax}
\end{equation}
Whenever $i_{\rm n}\sim 90^\circ$, the inner binary eccentricity approaches almost unity. If the octupole corrections are taken into account, the inner eccentricity can reach almost unity even if the initial inclination is outside of the $i_{\rm n}\sim 40^\circ$-$140^\circ$ KL range in the case the outer orbit is eccentric \citep{naoz13a,li14}.

The large values reached by the eccentricity of the BH-NS binary during the KL cycles make its merger time shorter since it efficiently dissipates energy when $e \sim e_{\rm in,n}^{\max}$ \citep[e.g., see][]{antognini14,fragrish2018}. However, KL oscillations can be suppressed by additional sources of precession, such as tidal bulges or relativistic precession. In our case, the most important effect is general relativistic precession that can quench the escursions to high eccentricity on a timescale \citep{nao16},
\begin{equation}
T_{GR}=\frac{a_{\rm in,n}^{5/2}c^2(1-e_{\rm in,n}^2)}{3G^{3/2}(\mbh+\mns)^{3/2}}\ .
\end{equation}
In the region of parameter space where $T_{KL}>T_{GR}$, the KL cycles of the BH-NS orbital elements are damped by relativistic effects.

\section{N-body simulations}
\label{sect:results}

\subsection{Initial conditions}

The stellar triples in our simulations are initialized as described in what follows. In total, we consider six different models (see Table~\ref{tab:models}).

For simplicity, we assume that every star in the mass range $8 \msun$--$20\msun$ will form a NS, while stars in the mass range $20 \msun$--$150\msun$ collapse to a BH. In all our models, we sample the mass $m_1$ of the most massive star in the inner binary from an initial mass function
\begin{equation}
\frac{dN}{dm} \propto m^{-\beta}\ ,
\label{eqn:bhmassfunc}
\end{equation}
in the mass range $20\msun$-$150\msun$, reflecting the progenitor of the BH. In our \textit{fiducial} model, $\beta=2.3$ (canonical \citet{kroupa2001} mass function; first model in Table~\ref{tab:models}). We adopt a flat mass ratio distribution for both the inner binary, $m_2/m_1$, and the outer binary, $m_3/(m_1+m_2)$. This is consistent with observations of massive binary stars, which suggest a nearly flat distribution of the mass ratio \citep{sana12,duch2013,sana2017}. The mass of the secondary in the inner binary is sampled from the range $8\msun$-$20\msun$, thus it is the progenitor of the NS, while the mass of the third companion is drawn from the range $0.5\msun$-$150\msun$.

The distribution of the inner and outer semi-major axis, $\ain$ and $\aout$, respectively, is assumed to be flat in log-space (\"{O}pik's law), roughly consistent with the results of \citet{kob2014}. We set as minimum separation $10$ AU, and adopt different values for the maximum separation $\amax=2000$ AU--$3000$ AU--$5000$ AU \citep{sana2014}. For the orbital eccentricities of the inner binary, $\ein$, and outer binary, $\eout$, we assume a flat distribution. We also run one additional model where we consider a thermal distribution of eccentricities, for comparison.

The initial mutual inclination $i_0$ between the inner and outer orbit is drawn from an isotropic distribution (i.e. uniform in $\cos i_0$). The other relevant angles are drawn randomly.

After sampling the relevant parameters, we check that the initial configuration satisfies the stability criterion of hierarchical triples of \citet{mar01},
\begin{equation}
\frac{R_{\rm p}}{a_{\rm in}}\geq 2.8 \left[\left(1+\frac{m_{\rm 3}}{m_1+m_2}\right)\frac{1+\eout}{\sqrt{1-\eout}} \right]^{2/5}\left(1.0-0.3\frac{i_0}{\pi}\right)\ ,
\label{eqn:stabts}
\end{equation}
where $R_p=\aout(1-\eout)$ is the pericentre distance of the outer orbit.

Given the above set of initial conditions, we let the primary star and the secondary star in the inner binary to undergo a conversion to a BH and NS, respectively. For simplicity, we assume that every star in the mass range $8 \msun$--$20\msun$ will form a NS of mass $\mns=1.3\msun$, while a star in the mass range $20 \msun$--$150\msun$ collapses to a BH of mass $\mbh=m/3$ \citep{sil17}, i.e. loses two-thirds of its mass. This is consistent with recent theoretical results on pulsational pair instability that limit the maximum BH mass to $\sim 50\msun$ \citep{bel2016b}. We note that the exact value of the mass intervals that lead to the formation of NSs and BHs and the mass lost during the MS phase and the following SN depends on the datailed physics of the star, importantly on the metallicity, stellar winds and rotation. These assumptions could affect the relative frequency of NSs and BHs, and the mass distribution of the latter ones. The orbital elements of the inner and outer orbit are updated as appropriate following the procedure discussed in Section~\ref{sect:supern}, to account both for mass loss and natal kicks.

The distribution of natal kick velocities of BHs and NSs is unknown. We implement momentum-conserving kicks, in which we assume that the momentum imparted to a BH is the same as the momentum given to a NS. As a consequence, the kick velocities for the BHs will be reduced with respect to those of NSs by a factor of $\mbh/\mns$. In our fiducial model, we consider a non-zero natal kick velocity for the newly formed BHs and NSs, by adopting Eq.~\ref{eqn:vkick} with $\sigma=\signs=260 \kms$. This is consistent with the distribution deduced by \citet{hobbs2005}. We run an additional model where we adopt $\sigma=\signs=100 \kms$, consistent with the distribution of natal kicks found by \citet{arz2002}. Finally, we adopt a model where no natal kick is imparted during BH and NS formation. For NSs, this would be consistent with the electron-capture supernova process, where the newly-formed NSs would receive no kick at birth or only a very small one due to an asymmetry in neutrino emissions \citep{pod2004}. We note that even in this case, the triple experiences a kick to its center of mass because one of the massive components suddenly loses mass \citep{bla1961}.

In case also the third companion is more massive than $8\msun$, we let it undergo an SN event and conversion to a compact object. If the triple remains bound, we check again the triple stability criterion of \citet{mar01} with the updated masses and orbital parameters for the inner and outer orbits.

Given the above set of initial parameters, we integrate the equations of motion of the 3-bodies
\begin{equation}
{\ddot{\textbf{r}}}_i=-G\sum\limits_{j\ne i}\frac{m_j(\textbf{r}_i-\textbf{r}_j)}{\left|\textbf{r}_i-\textbf{r}_j\right|^3}\ ,
\end{equation}
with $i=1$,$2$,$3$, by means of the \textsc{ARCHAIN} code \citep{mik06,mik08}, a fully regularized code able to model the evolution of binaries of arbitrary mass ratios and eccentricities with high accuracy and that includes PN corrections up to order PN2.5. We performed $1000$ simulations for each model in Table~\ref{tab:models}. We fix the maximum integration time as \citep{sil17},
\begin{equation}
T=\min \left(10^3 \times T_{\rm KL}, 10\ \gyr \right)\ ,
\label{eqn:tint}
\end{equation}
where $T_{\rm KL}$ is the triple KL timescales. In the case the third companion is not a compact object, i.e. $m_3\le 8\msun$, we set as the maximum timescale the minimum between Eq.~\ref{eqn:tint} and its MS lifetime, which is simply parametrised as \citep[e.g.][]{iben91,hurley00,maeder09},
\begin{equation}
\tau_{\rm MS} = \max(10\ (m/\msun)^{-2.5}\,{\rm Gyr}, 7\,{\rm Myr})\ .
\end{equation}
In this situation, we also check if the third star overflows its Roche lobe \citep{egg83}. In such a case, we stop the integration\footnote{We do not model the process that leads to the formation of a white dwarf. If the tertiary becomes a white dwarf and the system remains bound, some of the systems could still merge via KL oscillations.}.

\subsection{Results}

\begin{figure} 
\centering
\includegraphics[scale=0.55]{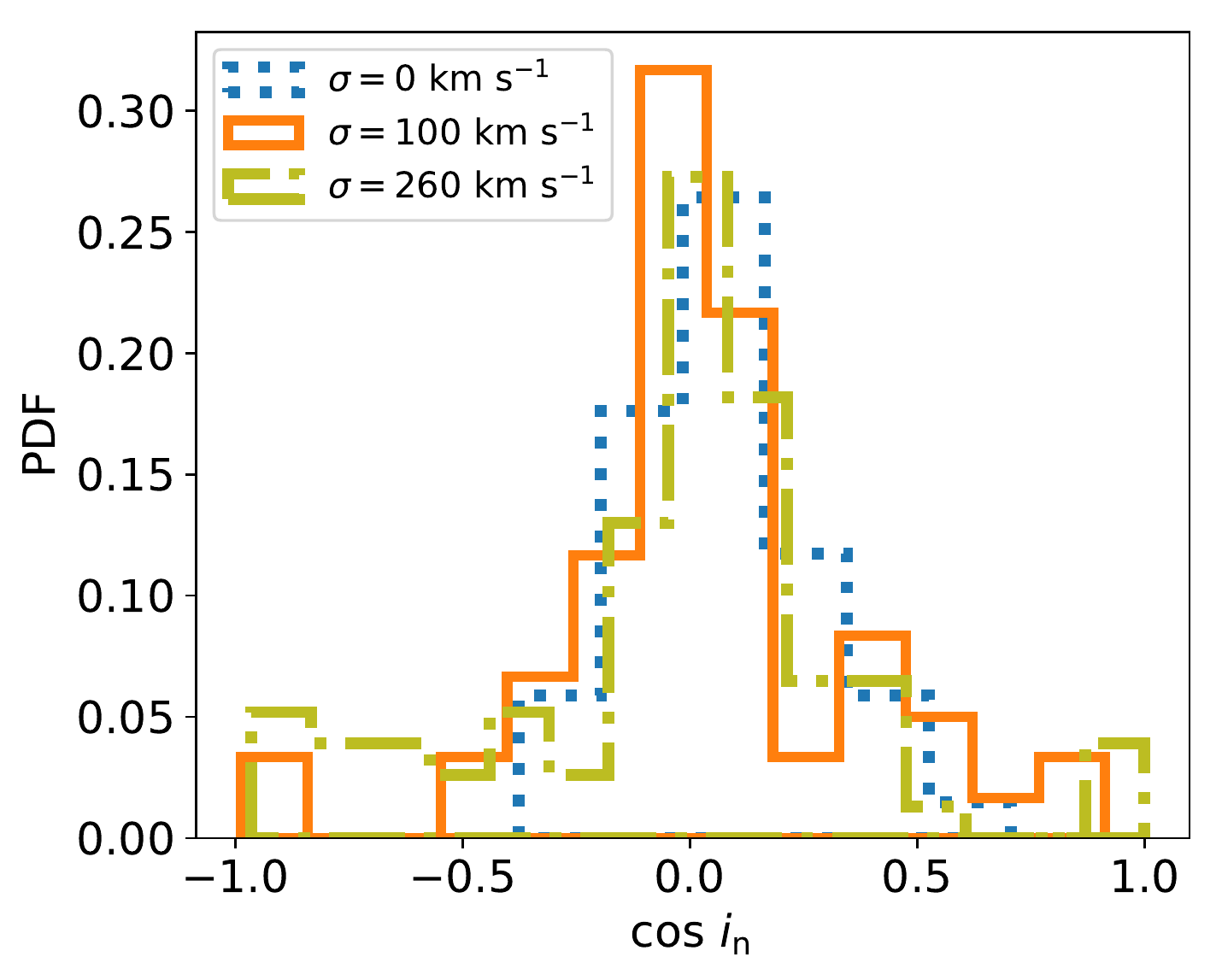}
\caption{Inclination ($i_{\rm n}$) distribution of BH-NS binaries that merge in triples. Most of the triples merge when $i_{\rm n}\sim 90^\circ$, where the effect of the KL oscillations is maximal.}
\label{fig:incl}
\end{figure}

A BH-NS binary is expected to be significantly perturbed by the tidal field of the third companion whenever its orbital plane is sufficiently inclined with respect to the outer orbit \citep{lid62,koz62}. According to Eq.~\ref{eqn:emax}, the BH-NS eccentricity reaches almost unity when $i_{\rm n}\sim 90^\circ$. Figure~\ref{fig:incl} shows the probability distribution function (PDF) of the initial binary plane inclination angle $i_{\rm n}$ (relative inclination after all the SN events in the triple take place (see Sect.~\ref{sect:supern})) in the systems which ended up in a merger. The distributions are shown for $\amax=2000$ AU and different values of $\sigma$. Independently of the mean of the natal kick velocity, the majority of the BH-NS mergers in triples takes place when the inclination approaches $\sim 90^\circ$. In this case, the KL effect is maximal, leading to eccentricity oscillates up to unity, the BH-NS binaries experience efficient gravitational energy dissipation near the pericentre and ends in a merger. BH-NS systems that merge with low relative inclinations have typically initial large eccentricities. 

\begin{figure} 
\centering
\includegraphics[scale=0.55]{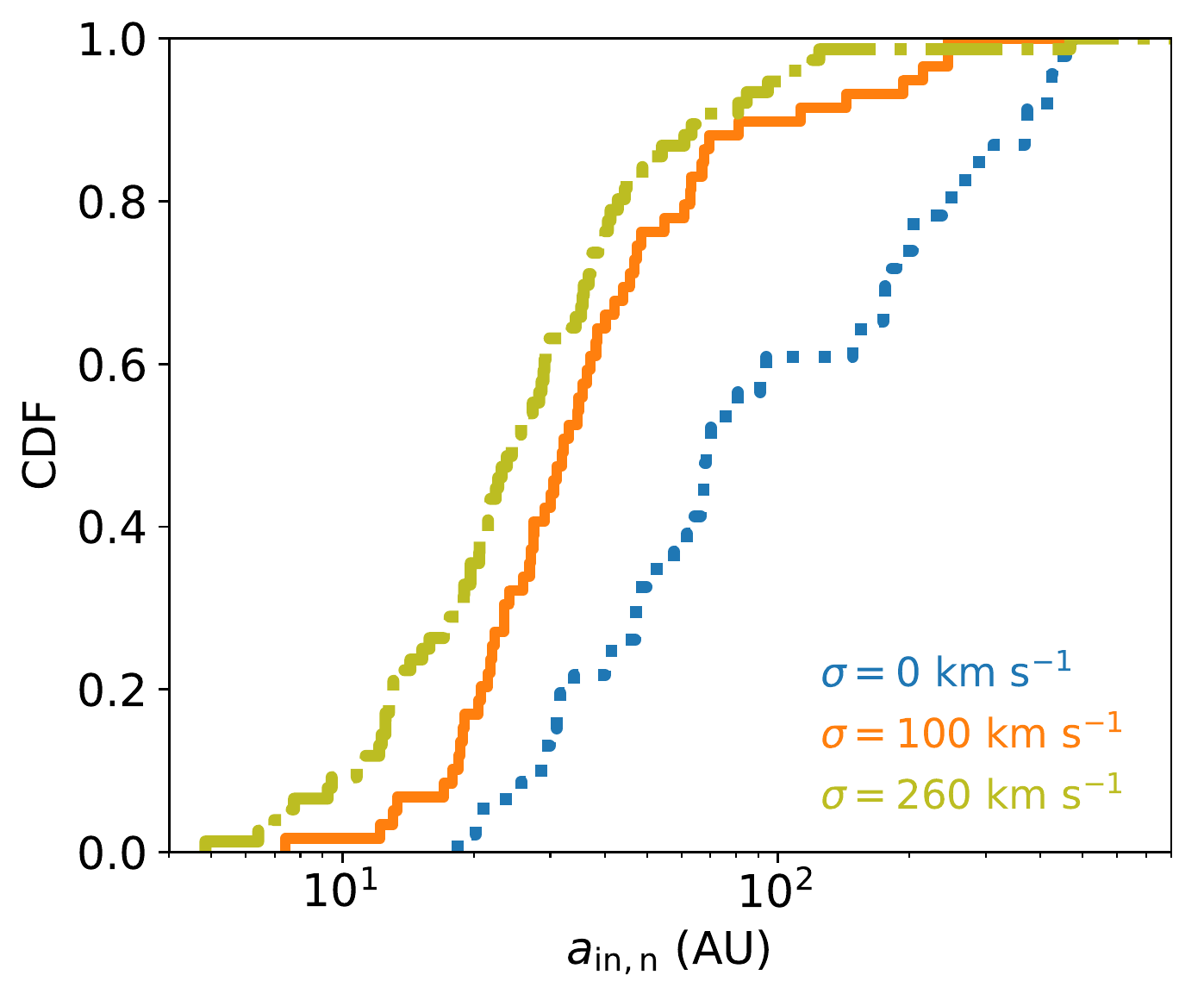}
\includegraphics[scale=0.55]{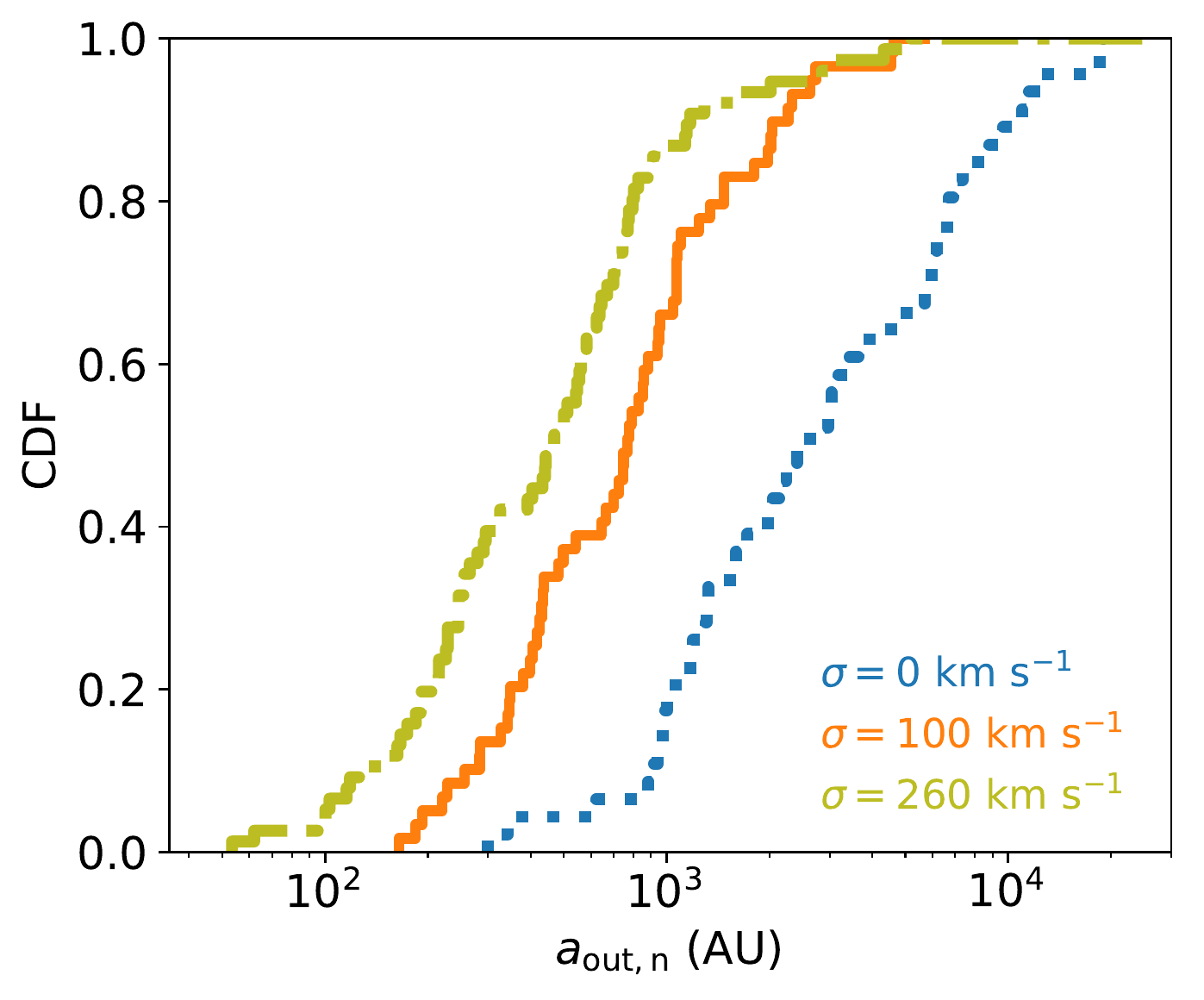}
\caption{Cumulative distribution function of inner (top) and outer (bottom) semi-major axis of BH-NS binaries in triples that lead to a merger, for different values of $\sigma$.}
\label{fig:semi}
\end{figure}

Figure~\ref{fig:semi} illustrates the cumulative distribution function (CDF) of inner (top) and outer (bottom) semi-major axis of BH-NS binaries in triples that lead to a merger, for different values of $\sigma$. In this figure, we plot $a_{\rm in,n}$ and $a_{\rm out,n}$, which are the inner and outer semi-major axes, respectively, after the SN events in the triple take place (see Sect.~\ref{sect:supern}). The larger the mean natal kick, the smaller the typical semi-major axis. This can be explained by considering that triples with wide orbits are generally unbound by large kick velocities, while they stay bound if the natal kick is not too high. We find that $\sim 50$\% of the systems have $a_{\rm in,n}\lesssim 70$ AU, $\lesssim 35$ AU, $\lesssim 25$ AU for $\sigma=0\kms$, $100\kms$, $260\kms$, respectively, and $\sim 50$\% of the systems have $a_{\rm out,n}\lesssim 2500$ AU, $\lesssim 800$ AU, $\lesssim 500$ AU for $\sigma=0\kms$, $100\kms$, $260\kms$, respectively. Since $T_{\rm KL}\propto a_{\rm out,n}^3/a_{\rm in,n}^{3/2}$, BH-NS systems that merge in models with $\sigma=0\kms$ are expected to typically merger on longer timescales compared to systems with non-null kick velocity, when the KL effect is at work (see Fig.~\ref{fig:tmerge}).  

\begin{figure} 
\centering
\includegraphics[scale=0.55]{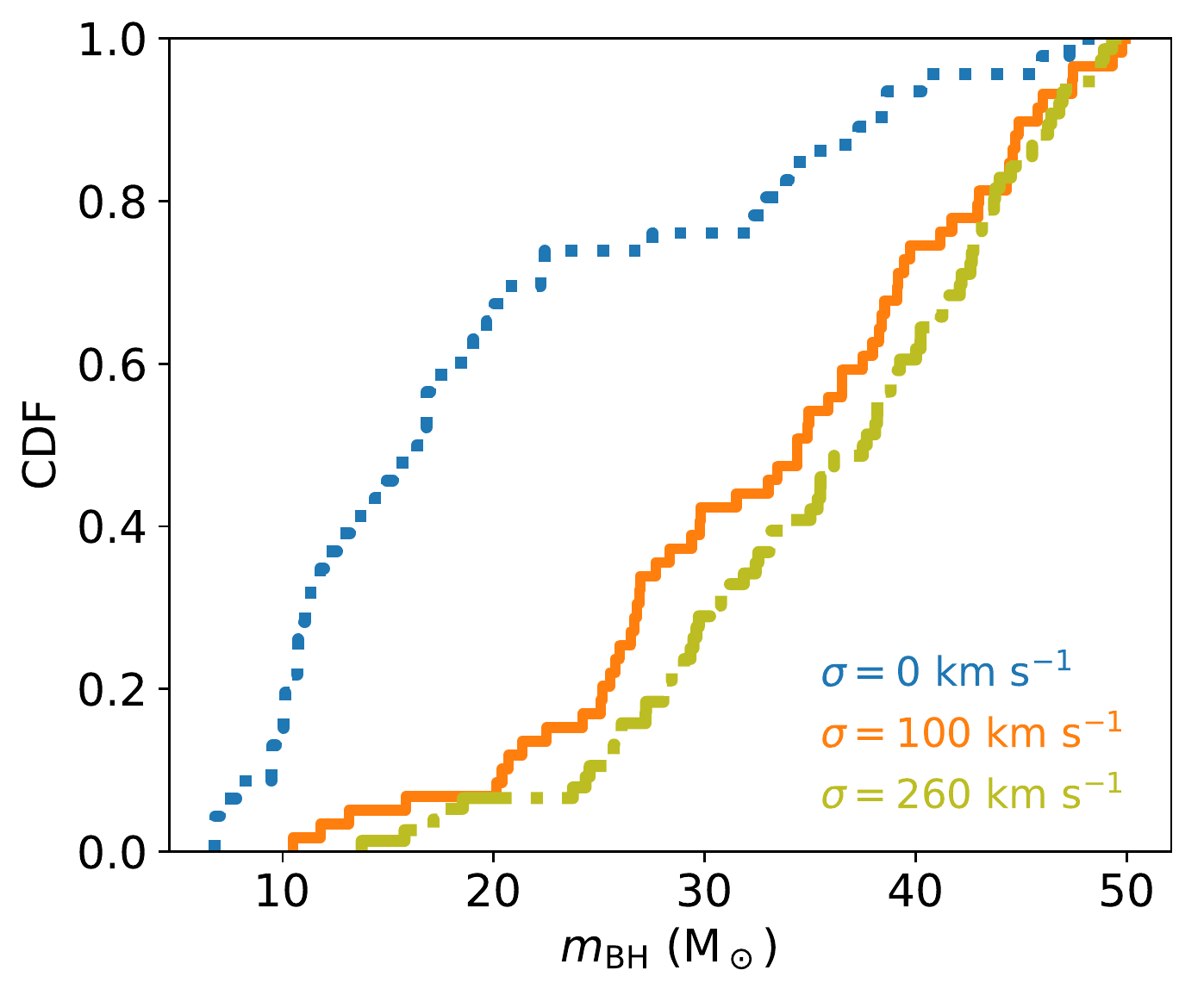}
\includegraphics[scale=0.55]{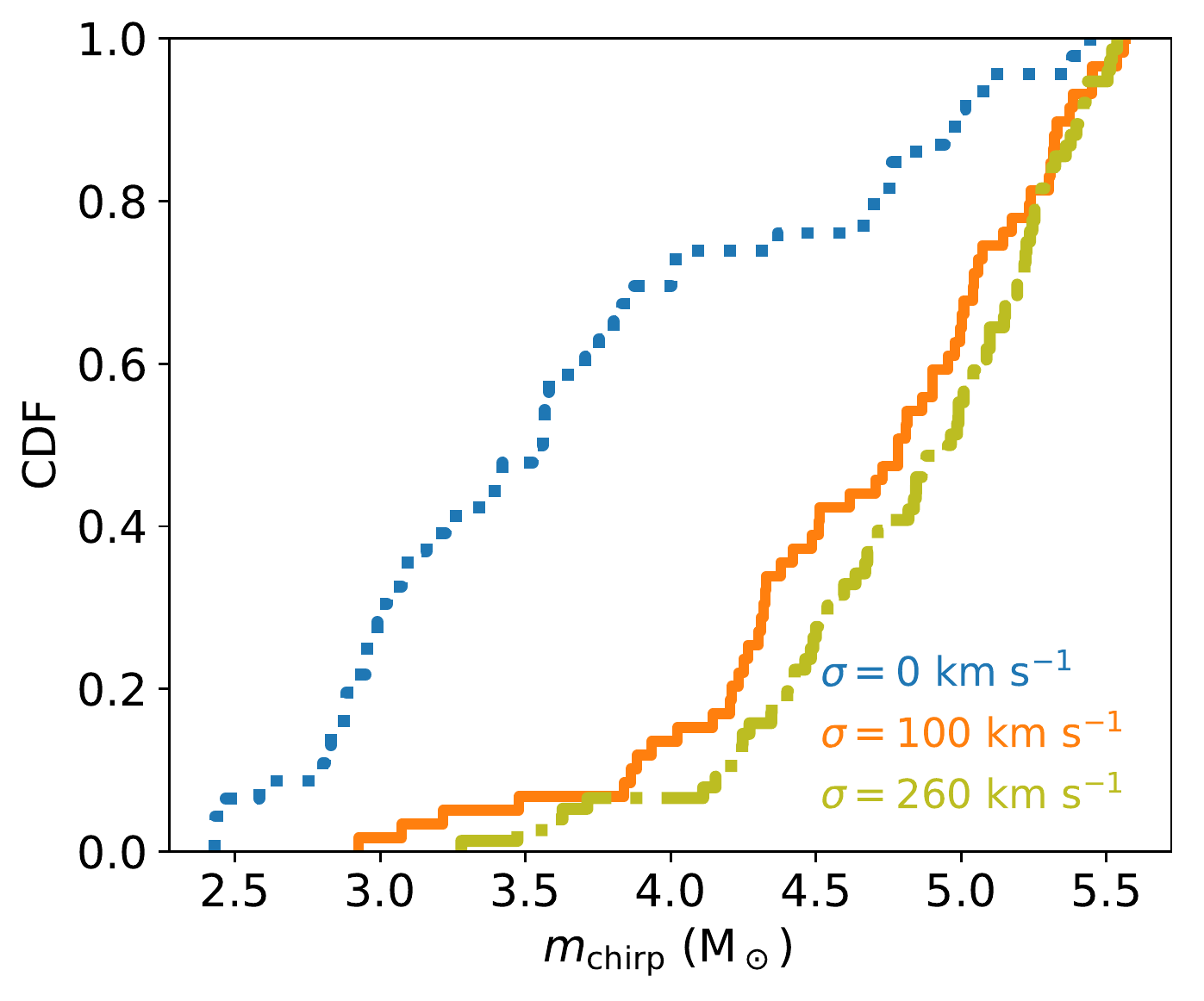}
\caption{Cumulative distribution function of total mass (top) and chirp mass (bottom) of BH-NS binaries in triples that lead to a merger, for different values of $\sigma$.}
\label{fig:mass}
\end{figure}

\begin{figure} 
\centering
\includegraphics[scale=0.55]{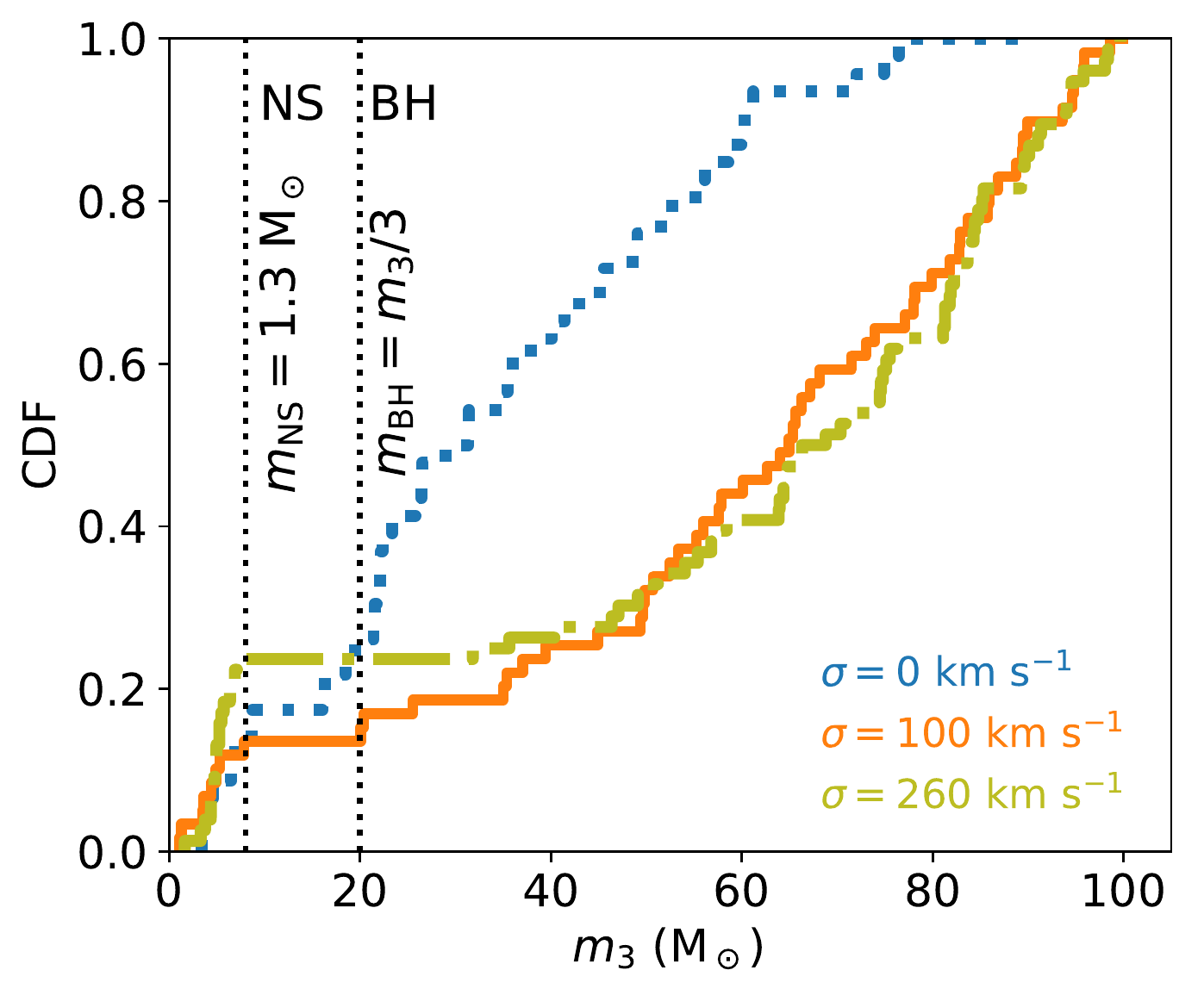}
\caption{Cumulative distribution function of the tertiary mass before it undergoes a SN (if any) of the BH-NS binaries in triples that lead to a merger, for different values of $\sigma$.}
\label{fig:mass3}
\end{figure}

The typical mean natal kick velocity affects also the distribution of BH masses in BH-NS binaries that merge in triple systems. We illustrate this in Figure~\ref{fig:mass}, where we plot the cumulative distribution function of total mass (top) and chirp mass (bottom) of BH-NS binaries in triples that lead to a merger. In the case of $\sigma=0\kms$, we find that merging BHs have typically lower masses compared to the models with $\sigma=100\kms$ and $\sigma=260\kms$. In the former case, $\sim 50\%$ of the BHs that merge have mass $\lesssim 17 \msun$, while for non-zero kick velocities we find that $\sim 50\%$ of the BHs that merge have mass $\lesssim 35 \msun$. This can be explained by our assumption of momentum-conserving kicks, where higher mass BHs receive, on average, lower velocity kicks and, as a consequence, are more likely to be retained in triples and (eventually) merge. However, other models for the kicks (e.g. scaling with the fraction of matter ejected by the SN event) may lead to different mass distributions. The same holds for the mass of the third companion. In Figure~\ref{fig:mass3}, we show the cumulative distribution function of the tertiary mass before it undergoes a SN (if any) of the BH-NS binaries in triples that lead to a merger, for different values of $\sigma$. All the third companions have initial mass $m_3\lesssim 100\msun$. Models with $\sigma=100\kms$ and $\sigma=260\kms$ typically predict a more massive tertiary, that will collapse to a BH of mass $m_3/3$, than the case of no natal kicks. Only in the Model where $\sigma=0\kms$, a few tertiaries are NSs. The fraction of third companions that will not collapse to a compact object is $\lesssim 20\%$.

\begin{figure} 
\centering
\includegraphics[scale=0.55]{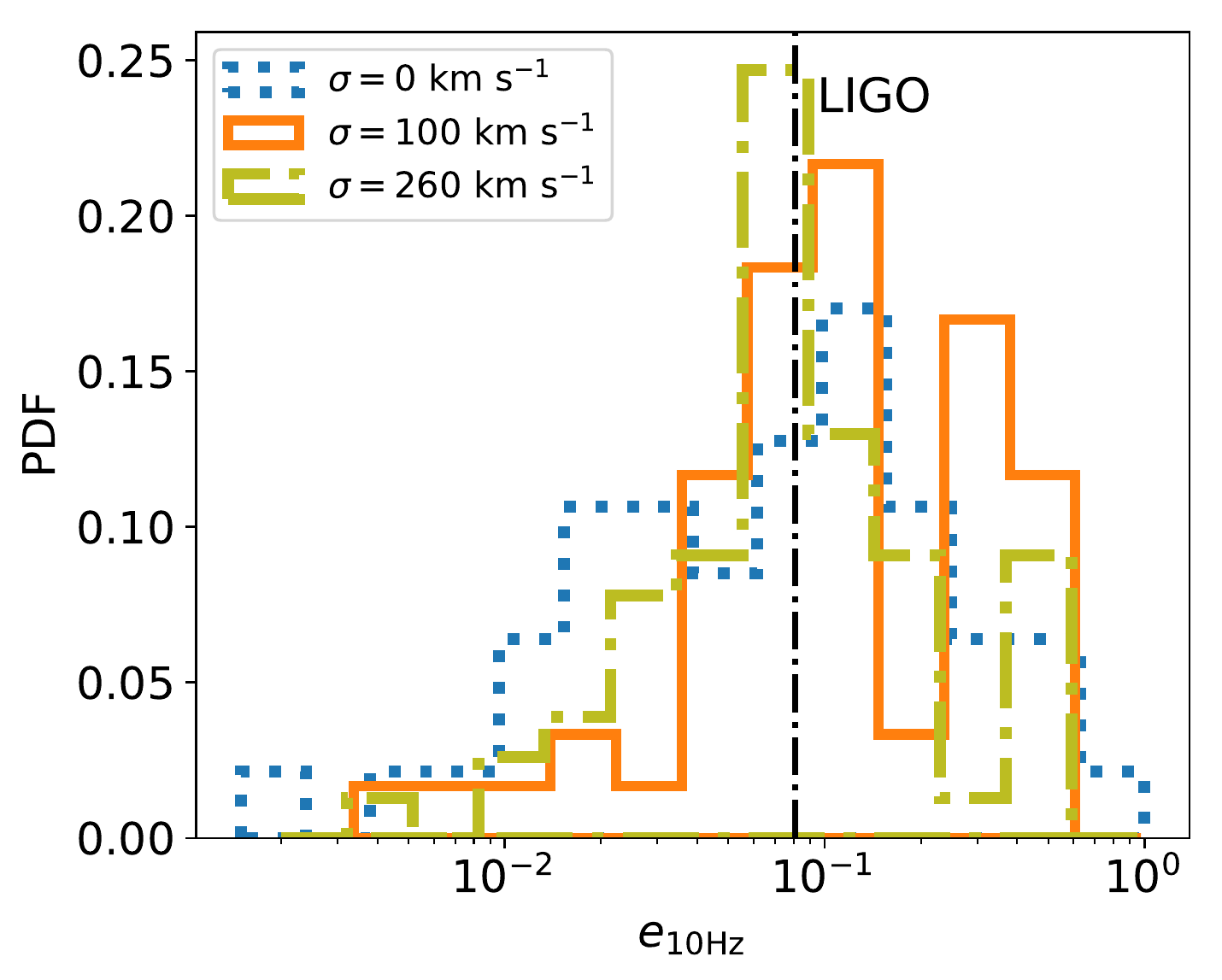}
\includegraphics[scale=0.55]{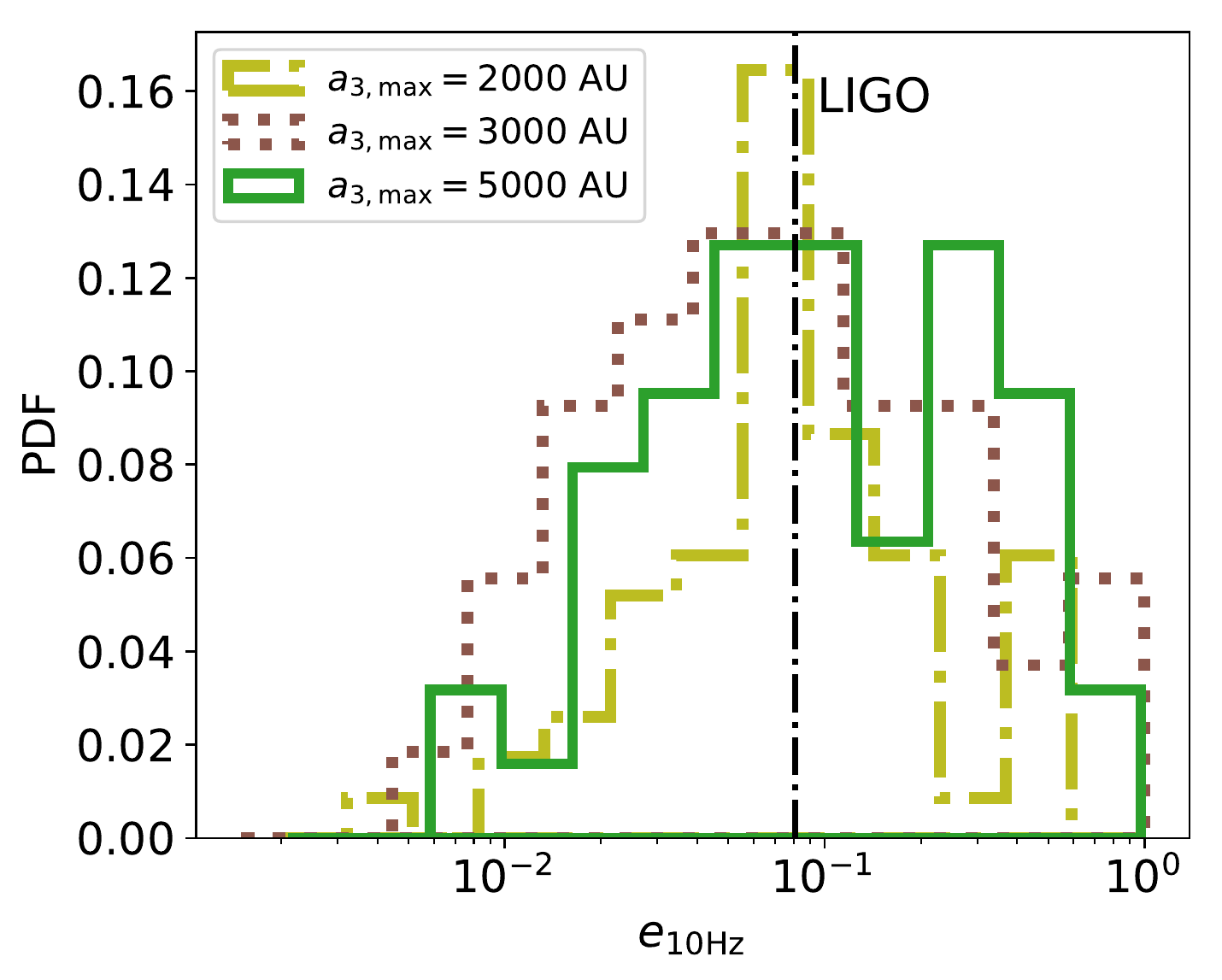}
\caption{Distribution of eccentricities at the moment the BH-NS binaries enter the LIGO frequency band ($10$ Hz) for mergers produced by triples. Top panel: $\amax=2000$ AU and different $\sigma$'s; bottom panel: $\sigma=260\kms$ and different $\amax$'s. The vertical line shows the minimum $e_{\rm 10Hz}=0.081$ where LIGO/VIRGO/KAGRA network may distinguish eccentric sources from circular sources \citep{gond2019}. A significant fraction of binaries formed in triples have a significant eccentricity in the LIGO band.}
\label{fig:ecc}
\end{figure}

Hierarchical configurations are expected to have eccentricities in the LIGO band ($10$ Hz) larger than binaries that merge in isolation, as a consequence of the perturbation of the third object and the KL cycles \citep[see e.g.][]{fragrish2018}. For the BH-NS binaries that merge in our simulations, we compute a proxy for the GW frequency, which we take to be the frequency corresponding to the harmonic that gives the maximal emission of GWs \citep{wen03},
\begin{equation} 
f_{\rm GW}=\frac{\sqrt{G(m_1+m_2)}}{\pi}\frac{(1+e_{\rm in,n})^{1.1954}}{[a_{\rm in,n}(1-e_{\rm in,n}^2)]^{1.5}}\ .
\end{equation}
In Figure~\ref{fig:ecc}, we illustrate the distribution of eccentricities at the moment the BH-NS binaries enter the LIGO frequency band for mergers produced by triples, for $\amax=2000$ AU and different $\sigma$'s (top panel) and $\sigma=260\kms$ and different $\amax$'s (bottom panel). We also plot the minimum $e_{\rm 10Hz}=0.081$ where LIGO/VIRGO/KAGRA network may distinguish eccentric sources from circular sources \citep{gond2019}. As expected for hierarchical configurations, a large fraction of BH-NS binaries formed in triple have a significant eccentricity in the LIGO band. On the other hand, BH-NS binaries that merge in isolation are essentially circular ($e\sim 10^{-8}$-$10^{-7}$) when they enter the LIGO frequency band. Thus, highly-eccentric mergers might be an imprint of BH-NS that merge through this channel. A similar signature could be found in BH-NS that merge in galactic nuclei in proximity to a supermassive black hole \citep{fragrish2018}, in mergers that follow from the GW capture scenario in clusters \citep{sam2018,sam2018b}, from hierarchical triples \citep{ant17} and quadruples \citep{fragk2019}, and from BH binaries orbiting intermediate-mass black holes in star clusters \citep{fragbr2019}

\subsection{Merger times and rates}

\begin{figure} 
\centering
\includegraphics[scale=0.55]{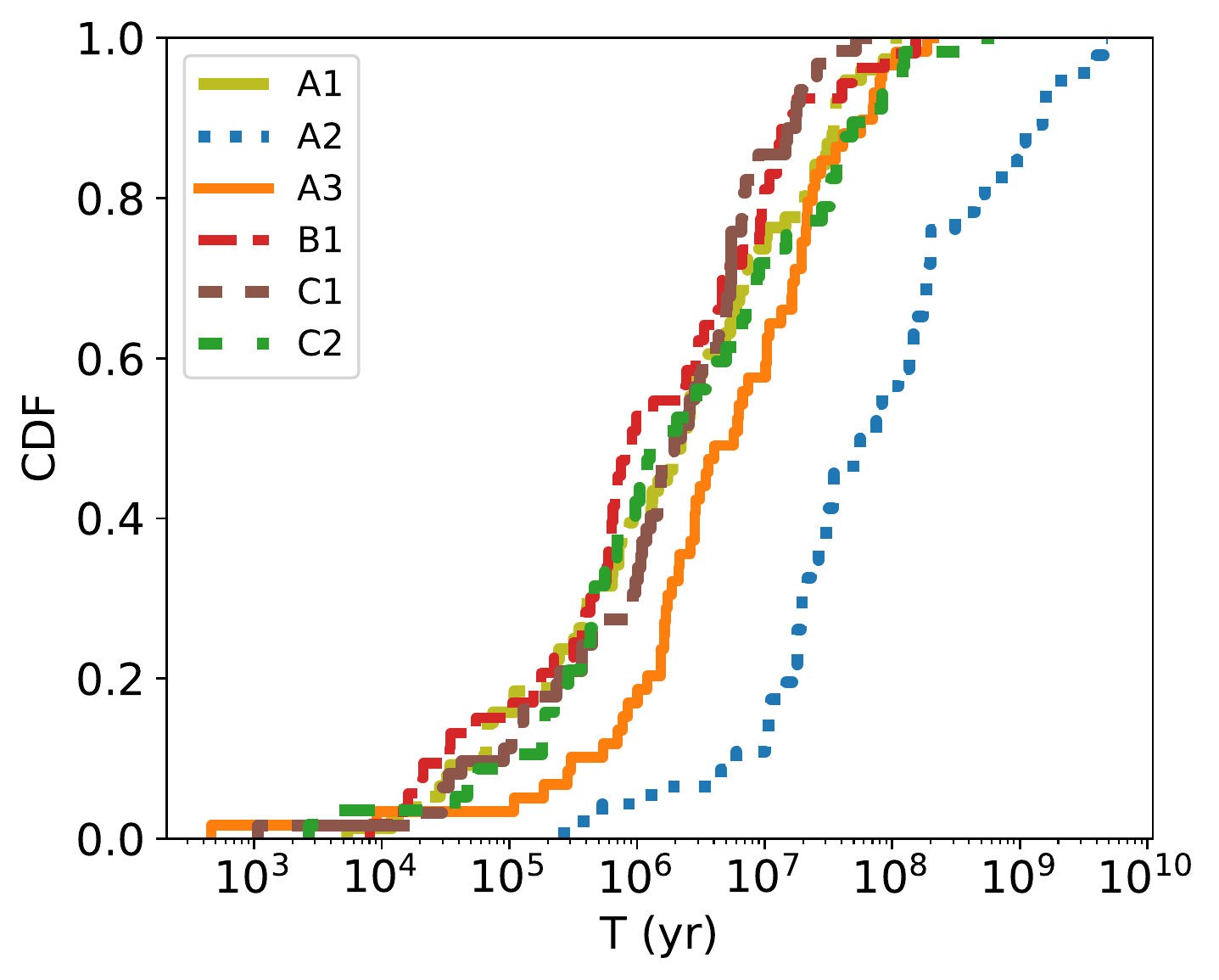}
\caption{Merger time distribution of BH-NS binaries in triples that lead to merger for all models (see Table\ref{tab:models}).}
\label{fig:tmerge}
\end{figure}

Figure~\ref{fig:tmerge} shows the merger time cumulative distribution functions of BH-NS binaries in triples that lead to merger for all models. The cumulative function depends essentially on the $\sigma$ of the natal velocity kick distribution. Larger kick velocities imply a larger outer semi-major axis, thus a larger KL timescale since $T_{\rm}\propto a_{\rm out,n}^{3}$. We find that BH-NS systems in models with $\sigma=0\kms$ merge on longer timescales compared to systems with non-null kick velocity, when the KL effect is at work. In our simulations, $\sim 50\%$ of the mergers happen within $\sim 2\times 10^6$ yr, $\sim 6\times 10^6$ yr, $\sim 8\times 10^7$ yr for $\sigma=0\kms$, $100\kms$, $260\kms$, respectively. Different $\amax$'s and a thermal distribution of the inner and outer eccentricity do not affect the merger time distribution significantly.

In order to compute the merger rate of BH-NS binaries, we assume that the local star formation rate is $0.025 \msun$ Mpc$^{-3}$ yr$^{-1}$, thus the number of stars formed per unit mass is given by \citep{both2011},
\begin{equation}
N(m)dm=5.4\times 10^6 m^{-2.3}\ \mathrm{Gpc}^{-3}\ \mathrm{yr}^{-1}\ .
\end{equation}
Adopting a constant star-formation rate per comoving volume unit, the merger rate of binary BH-NS in triples is then,
\begin{equation}
\Gamma_\mathrm{BH-NS}=8.1\times 10^4 f_{\rm 3} f_{\rm stable} f_{\rm merge}\ \mathrm{Gpc}^{-3}\ \mathrm{yr}^{-1}\ ,
\end{equation}
where $f_{\rm 3}$ is the fraction massive stars in triples, $f_{\rm stable}$ is the fraction of sampled systems that are stable after the SN events take place, and $f_{\rm merge}$ is the conditional probability that systems (stable after all the SNe) merge. In our calculations, we adopt $f_{\rm 3}=0.15$. The fraction of stable systems after SNe depends mainly on the value of $\sigma$ for the natal velocity kick distribution. We find $f_{\rm stable}\approx 1.6\times 10^{-2}$, $3.6\times 10^{-5}$, $1.1\times 10^{-6}$ for $\sigma=0\kms$, $100\kms$, $260\kms$, respectively, when $\amax=2000$ AU, and $f_{\rm stable}\approx 1.1\times 10^{-6}$, $0.9\times 10^{-6}$, $0.7\times 10^{-6}$ for $\amax=2000$ AU, $3000$ AU, $5000$ AU, respectively, when $\sigma=260\kms$. The typical fraction of systems that merge is $f_{\rm merge}=0.1$ (see Tab.~\ref{tab:models}). Therefore, our final estimated rate is in the range,
\begin{equation}
\Gamma_\mathrm{BH-NS}=1\times 10^{-3}-19 \ \mathrm{Gpc}^{-3}\ \mathrm{yr}^{-1}\ .
\end{equation}
Our rate overlaps with BH-NS mergers in binaries \citep{giac2018,kruc2018} and is entirely within the LIGO allowed values \citep{ligo2018}.

\section{Discussion and conclusions}
\label{sect:conc}

While BH-NS mergers have not been found yet by current GW observatories, many of these systems are expected to be observed in the upcoming years. LIGO has set a $90\%$ upper limit of $610$ Gpc$^{-3}$ yr$^{-1}$ for the merger rate. The formation of BH-NS binaries is particularly difficult in star clusters since the strong heating by BH interactions prevent the NSs from efficiently forming these binaries \citep{frag2018,ye2019}. This favours binary evolution in isolation \citep{giac2018,kruc2018}. However, surveys of massive stars have shown that a significant fraction of them resides in triple systems \citep{duq91,sana2017}.

We have carried out for the first time a systematic statistical study of the dynamical evolution of triples comprised of an inner BH-NS binary by means of direct high-precision $N$-body simulations, including Post-Newtonian (PN) terms up to 2.5PN order. In our calculations, we have started from the MS progenitors of the compact objects and have modeled the SN events that lead to the formation of BHs and NSs. Our findings are as follows:
\begin{itemize}
\item The majority of the BH-NS mergers in triples takes place when the inclination is $\sim 90^\circ$ and the KL effect is maximized, independently of the mean kick velocity.
\item $\sigma$ affects the distribution of orbital elements of the merging BH-NS binaries in triples; the larger the mean natal kick, the smaller the typical semi-major axis.
\item Larger $\sigma$'s lead to larger BH and chirp mass.
\item A large fraction of merging BH-NS binaries in triple have a significant eccentricity ($\gtrsim 0.1$) in the LIGO band.
\item Models with $\sigma=0\kms$ merge on longer timescales compared to systems with a natal kick velocity.
\end{itemize}

We have also computed the merger rate of BH-NS binaries, and found a rate
$\Gamma_\mathrm{BH-NS}=1\times 10^{-3}-19 \ \mathrm{Gpc}^{-3}\ \mathrm{yr}^{-1}$, which overlaps with the BH-NS mergers in binaries \citep{giac2018,kruc2018} and is entirely within the LIGO range of allowed values \citep{ligo2018}. The large uncertainty in the rate originates from the unknown physics and natal kick velocity distribution of compact objects in triple system. Therefore, the observed merger rate of BH-NS binaries can be used to constrain the magnitude of the velocity kicks that are imparted during the SN process. Future work is also deserved to the study of the different assumption on the final mass of the NSs and BHs. The exact value of the mass intervals that lead to the formation of NSs and BHs and the mass lost during the MS phase and the following SN depends on the stellar metallicity, winds and rotation. These assumptions could affect the relative frequency of NSs and BHs, and the BH mass distribution.

BH-NS mergers are also of interest for their electromagnetic (EM) counterparts, such as short gamma-ray bursts, which can provide crucial information on the origin of BH-NS mergers, since they can potentially provide a much better localization and redshift determination compared to GWs alone. Mergers of BH-NS binaries could be accompanied by the formation of an hyperaccreting disk whenever the mass ratio does not exceed $\sim 5$--$6$ \citep{Pannarale2011,Foucart2012,Foucart2018}. We find that $\sim 5$\% of mergers for $\sigma=0\kms$ have mass ratios $\lesssim 5$--$6$. For $\sigma=100\kms$ and $\sigma=260\kms$ we find no such systems. No disk will form for larger mass ratios since the tidal disruption radius of the NS is smaller than the radius of the innermost stable circular orbit, thus resulting in a direct plunge into the BH \citep{Bartos2013}. Larger mass ratios would instead reveal important information on the NS magnetosphere and crust \citep{tsang2012,dora2013,dora2016}.

\section*{Acknowledgements}

GF is supported by the Foreign Postdoctoral Fellowship Program of the Israel Academy of Sciences and Humanities. GF also acknowledges support from an Arskin postdoctoral fellowship. GF thanks Seppo Mikkola for helpful discussions on the use of the code \textsc{archain}. Simulations were run on the \textit{Astric} cluster at the Hebrew University of Jerusalem. This work was also supported by the Black Hole Initiative at Harvard University, which is founded by a JTF grant (to AL).

\bibliographystyle{mn2e}
\bibliography{refs}

\end{document}